\documentclass[pre,preprint,floatfix]{revtex4-1} 

\usepackage{amsmath} 
\usepackage{amsfonts} 
\usepackage{graphicx} 

\begin{document}


\title{Low frequency spectra of bending wave turbulence}
\author{Benjamin Miquel}
\affiliation{Université Paris-Saclay, CNRS, CEA, Service de Physique de l’État Condensé, 91191 Gif-sur-Yvette, France,}
\author{Antoine Naert}
\affiliation{Univ Lyon, ENS de Lyon, Univ Lyon 1, CNRS, Laboratoire de Physique, F-69342 Lyon,France }
\author{S\'ebastien Auma\^ \i tre}
\email[Corresponding author. Email address: ]{sebastien.aumaitre@cea.fr}
\affiliation{Univ Lyon, ENS de Lyon, Univ Lyon 1, CNRS, Laboratoire de Physique, F-69342 Lyon,France}
\affiliation{Université Paris-Saclay, CNRS, CEA, Service de Physique de l’État Condensé, 91191 Gif-sur-Yvette, France}

\begin{abstract}
{We study experimentally the dynamics of long waves among  turbulent bending waves in a  thin elastic plate set into vibration by a monochromatic forcing at a frequency $f_0$. This frequency is chosen large compared with the characteristic frequencies of bending waves. As a consequence, a range of conservative scales, without energy flux in average, exists for frequencies $f<f_0$. Within this range, we report a flat power density spectrum for the orthogonal velocity, corresponding to energy equipartition between modes. Thus, the average energy per mode $\beta^{-1}$---analogous to a temperature---fully characterizes the large-scale turbulent wave field. We present an expression for $\beta$ as a function of the forcing frequency and amplitude, and of the plate characteristics.}  

\end{abstract}
\maketitle 

{\bf Introduction}\\
Turbulence  generically refers to physical systems where a large number of degrees of freedom exchange energy through nonlinear interactions and exhibit chaotic dynamics. A salient feature of these out-of-equilibrium systems is the existence of a scale separation between the characteristic scales of energy injection and of energy dissipation. From a phenomelogical point of view, this scale separation permits energy cascades, a canonical example of which is the direct (i.e., down-scale) energy cascade in hydrodynamical turbulence (HT) between the (large) injection scale and the (small) dissipation scale \cite{Richardson,BiferaleAlexakis}. From a theoretical point of view, this scale separation advocates for a statistical approach~\cite{Yaglom,Frish}. For instance, the Kolmogorov 1941 theory~\cite{K41} (thereafter K41) captures successfully many statistical properties of the direct energy cascade in HT, though refined theories have been formulated later \cite{Yaglom,Frish,K62}. From a practical point of view, most numerical and experimental studies of hydrodynamic turbulence focus on this direct cascade and therefore consider fluid domains comparable to  the forcing scale. However, larger structures might exist, provided that the system size allows for it . They haive received much less attention, despite their relevance in many geophysical and astrophysical contexts \cite{BouchetVenaille}. A computationally-savvy approach consists in modelling the large scale modes only by considering the truncated Euler equation (e.g. \cite{FauveAlexakis,BrachetAlexakis}). These studies have recently established the generally accepted theoretical view that, in absence of an energy flux through scales larger than the forcing scale, energy is statistically evenly distributed between these large scales~\cite{Saffman,Kraichnan,Sulem}. This large scale energy equipartition is the hydrodynamic analogous to the thermal equilibrium. 

Energy cascades are not exclusive to hydrodynamical turbulence, but occur in other systems as well. For instance, cascades develop in a variety of nonlinear wave fields, which constitutes the focus of the present manuscript. Similarly to the hydrodynamical case, a statistical theory has been derived in the limit of weak nonlinearity: the weak wave turbulence theory (WWT)  \cite{ZakharovLvovFalkovich,NewellRumpf,Nazarenko}.  The most prominent success of the WWT is perhaps to derive analytically power-law spectra associated with a non-zero energy flux through scales $\epsilon$. These spectra, coined Kolmogorov-Zakharov spectra, often take the form of power-laws in both $k$ and $\epsilon$, the exponents of which depend on the nature of the interactions (dimensionality, number of interacting waves, dispersion relation, etc.). The theory has a wide range of applicability and has been successfully adapted to a variety of wave systems: gravity and capillary surface waves, sound waves, Alfven waves, plasma waves, internal waves, nonlinear optics, Bose-Einstein condensates and gravitational waves \cite{ZakharovLvovFalkovich,NewellRumpf,Nazarenko,Galtier}.  

Among nonlinear interacting waves, we consider here the bending waves propagating in a thin elastic plate \cite{Landau, Amabili,Audoly}. A turbulent spectrum has been proposed for the direct cascade in the WWT framework \cite{RicaJosserand}. The prediction differs from experimental observations \cite{Mordant,Cadot}. The discrepancy is attributed to the dissipation mechanisms \cite{Humbert,BenjaminPRE2014}. The nonlinear interactions of elastic bending waves involves four waves but, in contrast to surface gravity waves, the number of wave in interaction is not necessarily conserved \cite{RicaJosserand}. However, the existence of an inverse cascade (i.e. towards large scales) is still an open question. In reference \cite{DuringJosserandRica}, authors observed an inverse cascade growing up during transient regime. They attributed this transient inverse cascade to the interactions conserving the number of waves. In this numerical study, the direct cascade is shrunk to a very short range in order to extend the inverse cascade. Moreover, the final stationary state is unknown.

 Our aim in this article is to characterize experimentally the low frequency spectrum of nonlinear bending waves in the stationary regime. Similar studies were performed only recently on surface gravity waves where an inverse cascade exists \cite{Dysturb} and on capillary waves where no energy flux is expected to larger scales until the transition to gravity waves \cite{GuillaumePRL}. As we demonstrate below, flexion waves in thin steel plates prove a valuable candidate to study the large scale dynamics of turbulent systems, where one can easily excite and measure waves with temporal frequencies 1 Hz $\leq f \leq$ 300 Hz (slower than the forcing) and with wavelength 1 cm$\leq \lambda \leq$ 100cm.. 

{\bf Experimental Device } 

\begin{figure}
\centerline{\resizebox{0.75\textwidth}{!}{%
\includegraphics{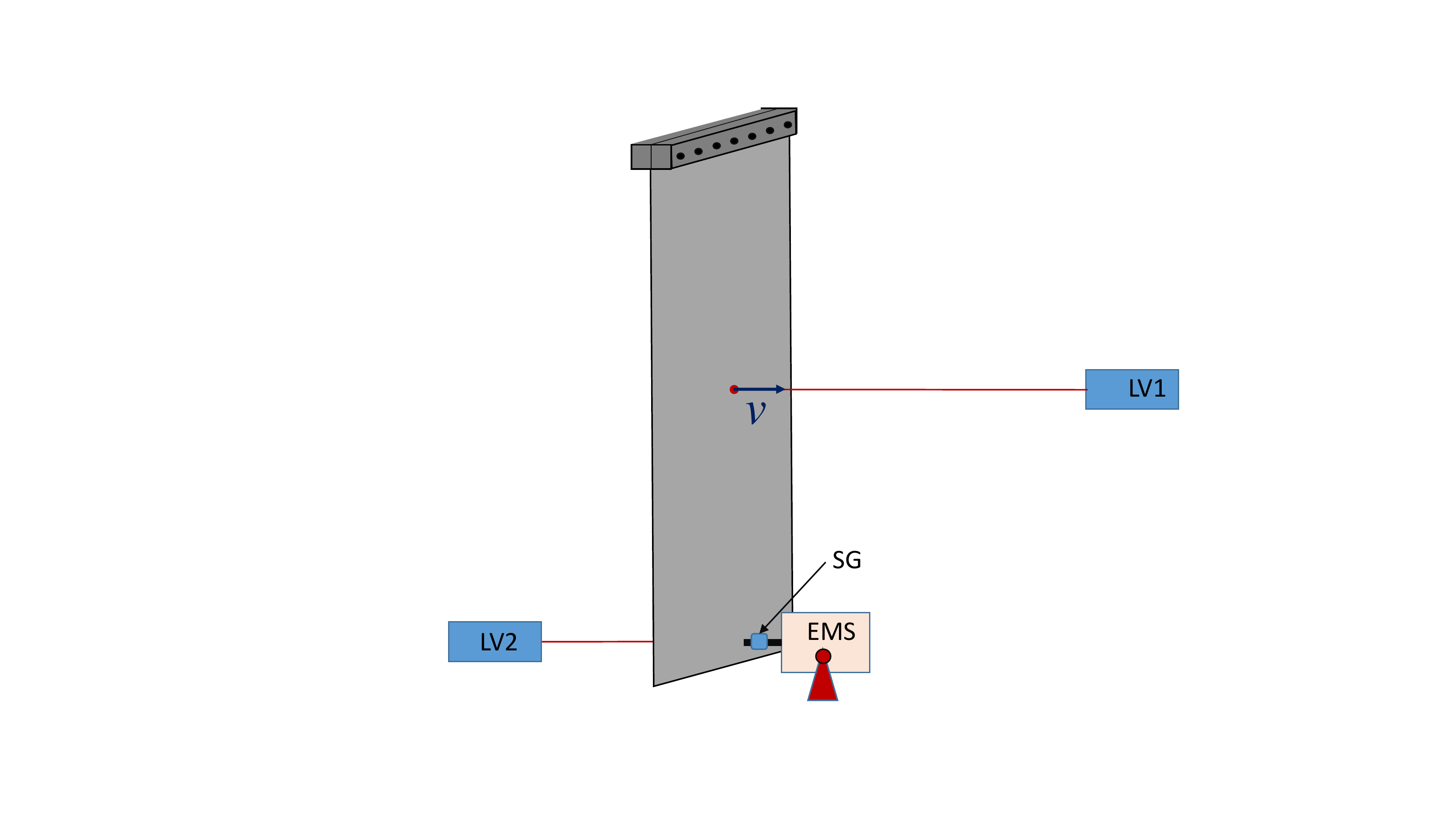}}}
\caption{Experimental setup: a thin stainless steel plate of ($2000\times1000\times 0.5$ mm$^3$) is forced by a electromagnetic shaker (EMS) at a frequency $f_o\in [100 -300]$ Hz. The Laser vibrometer (LV1) measures the velocity of the perpendicular deformation. In addition, we measure the force applied by the EMS with a strain gauge (SG) and the velocity at the injection point with a second laser vibrometer (LV2) [see text].}
\label{ExpDevLFrSp} 
\end{figure}
Our experimental device is similar to the one presented in \cite{AumaitreNaertAppfel}. It is sketched in figure \ref{ExpDevLFrSp}.  A thin stainless steel plate ($2000\times1000\times 0.5$ mm$^3$) is forced with an electromagnetic shaker (EMS). A harmonic forcing at frequency $f_0 \in[100-300]$ Hz is used. The forcing point is at 10 cm from the bottom edge and the top edge is clamped. All others boundaries are free. The low frequency cut-off, $f_c$, due to finite size of the plate, is estimated with the dispersion relation of the bending waves: $2\pi f=ch k^2$. Here $f$ and $k$ are respectively the bending wave frequency and wave number, $h$ is the plate thickness and $c$ is a constant (with the dimension of a velocity) depending only on the mechanical properties of the plate materials \cite{Landau}. In \cite{Cadot}, the authors estimate $c=1570$ m/s in a similar plate. Taking the cutoff wave number $k_c\sim1/L$ with $L=1$ m the plate width, one gets $f_c\sim 5$ Hz. In the following, we are interested in the spectra between $f_c$ and $f_o$ of the normal velocity $v$ corresponding to displacements along the direction perpendicular to the plate at rest. This velocity is measured locally with a laser vibrometer (LV1) near the center of the plate. We checked that the velocity spectrum does not depend on the exact position of the laser spot as long as it is 20 cm apart from the plate boundaries. A second laser vibrometer (LV2) measures the perpendicular velocity at the energy injection point where the EMS is attached to the plate. In addition, we measure the force applied by the EMS with a strain gauge sensor (SG). The mechanical power injected into the nonlinear bending wave field, denoted $I$, is inferred from the product of these two measurements. In the next section, we present the low frequency spectra of the perpendicular velocity $v$  for various injected powers and driving frequencies. For each frequency and amplitude, we record the plate velocity for a quarter of an hour in the stationary regime. We probe each set of forcing parameters four times. The  comparison of each four similar measurements gives us an estimate of the experimental reproducibility.

{\bf Experimental results}

\begin{figure}
\vspace{-3.cm}
\centerline{\resizebox{0.55\textwidth}{!}{%
\includegraphics{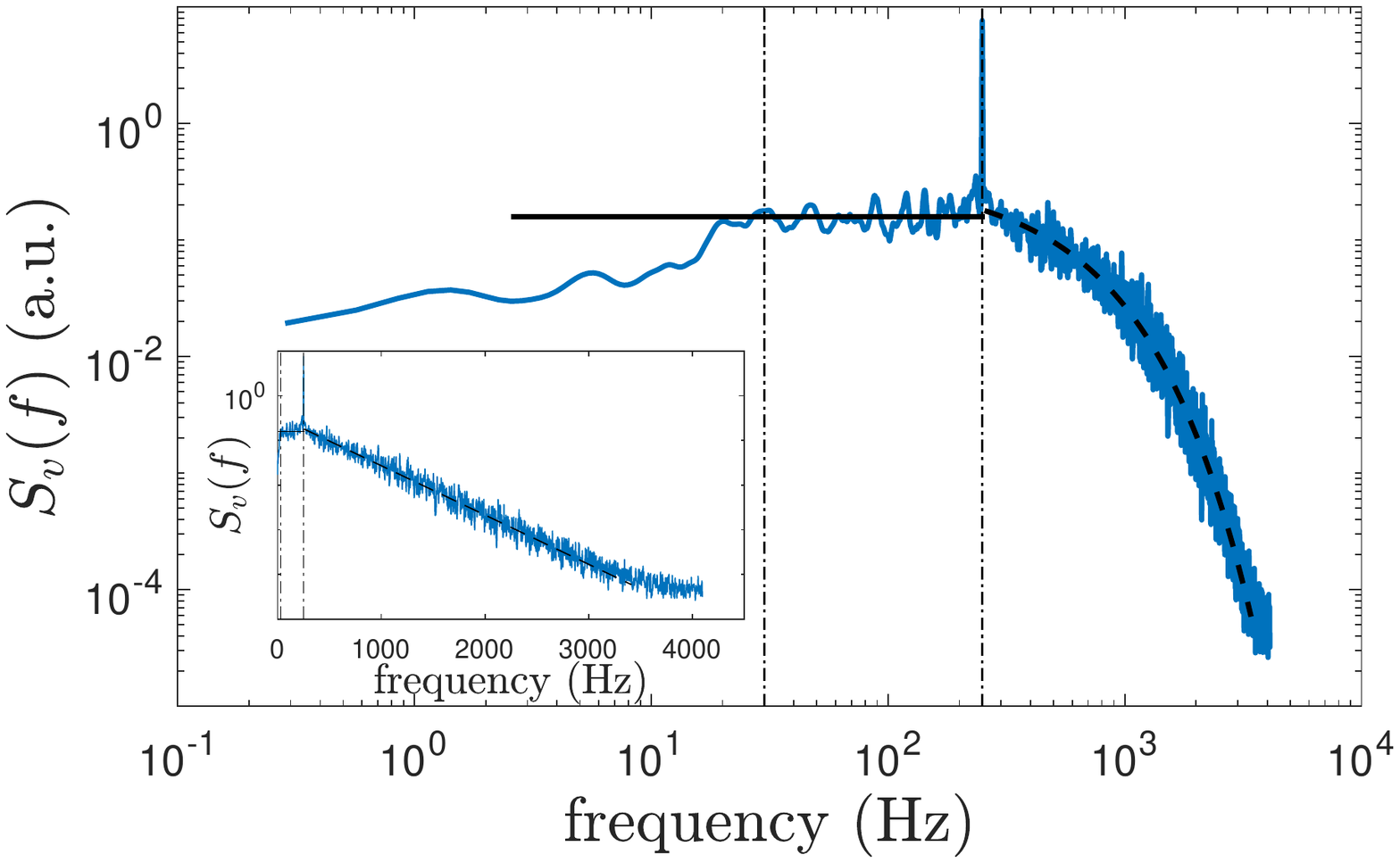}}

}
\vspace{-4.cm}
\caption{PDS of the perpendicular velocity measured at a point in the middle of the plate with $f_o=250$\,Hz and an input power of 407 mW. The continuous line represents the plateau value evaluated between 30 and 220 Hz. {\bf Main panel}: the PDS is plotted with logarithmic axes. {\bf Inset}: semi-logarithmic axes are used. The vertical dot-dashed line point out the low frequency cutoff at $f_c=30$ Hz and the forcing frequency at $f_o=250$ Hz. }
\label{PSDv300} 
\end{figure}

\begin{figure}
\vspace{-3.cm}
\centerline{\resizebox{0.55\textwidth}{!}{%
\includegraphics{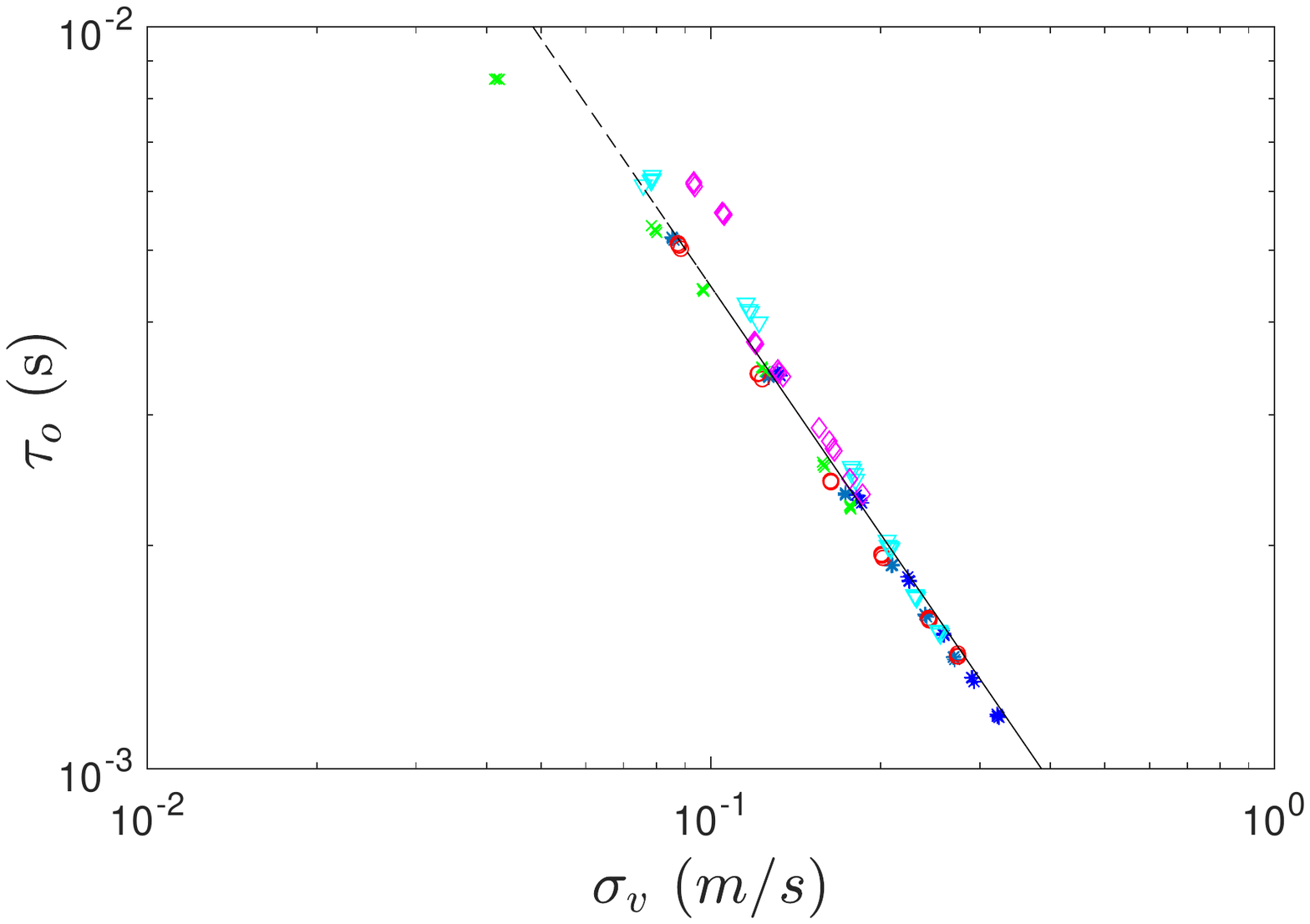}}
}
\vspace{-4cm}
\caption{ The characteristic time of the exponential decay of the PDS of $v$, as a function of the standard deviation of $v$. The various forcing frequencies are: $f_o=100$ Hz  $*$(blue online), $f_o=150$ Hz $\circ$ (red) , $f_o=200$ Hz $\times$ (green) , $f_o=250$ Hz $\diamond$ (magenta), $f_o=300$ Hz $\nabla$ (cyan). The dashed line  represents the best fit of the experimetal data. It gives a exponent of -1.1$\pm$0.1.}
\label{tauvsv} 
\end{figure}

Figure \ref{PSDv300} presents the Power Density Spectrum (PDS) of the perpendicular velocity, $S_{v}$ for a forcing frequency of $f_o=$250 Hz and an input power of 407 mW.  The forcing frequency remains clearly visible in the spectrum. Despite its amplitude, this peak actually contains only few percent of the total energy due to its narrowness. At frequencies higher than $f_o$, data are accurately fitted with an exponential decay $C(I,f_o)\exp(-\tau_o f)$, shown as the dashed line in the inset of figure \ref{PSDv300}. A similar exponential decay is found in the viscous range of hydrodynamic turbulence spectra~\cite{Yaglom,Kraichnan59}. Here $\tau_o$ is a characteristic decaying time and $C(I,f_o)$ is a parameter depending on the forcing. This kind of decay holds for the entire range of driving parameters ($I$ and $f_o$) explored in our experiment. Moreover, figure \ref{tauvsv} shows that the characteristic time $\tau_o$ is, to a good approximation, inversely proportional to the rms velocity and independent of the forcing frequency, $f_o$, i.e $\tau_o=\alpha/\sigma_v$ where $\alpha$ is a constant length independent of the forcing. From figure \ref{tauvsv} one estimates: $\alpha\simeq0.35$ mm. 

\begin{figure}
\vspace{-4.cm}
\centerline{\resizebox{0.55\textwidth}{!}{%
\includegraphics{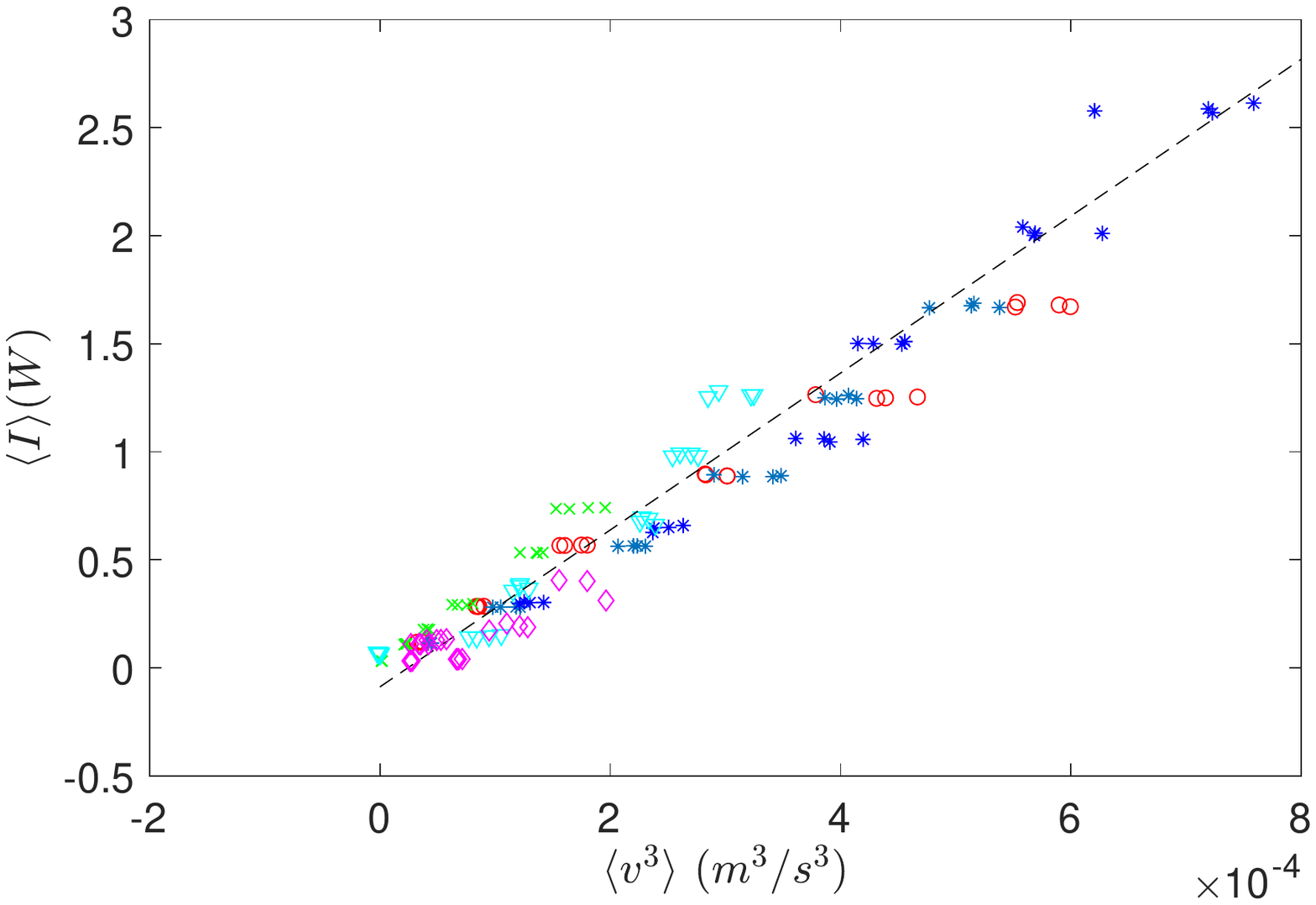}}
}
\vspace{-4.cm}
\caption{The mean injected power $\langle I \rangle$ as a function of the 3rd moment of the perpendicular velocity $v$ . Symbols are as previously:  $f_o=100$ Hz  $*$(blue online), $f_o=150$ Hz $\circ$ (red) , $f_o=200$ Hz $\times$ (green) , $f_o=250$ Hz $\diamond$ (magenta), $f_o=300$ Hz $\nabla$ (cyan). The dashed line represents the best linear fit wit a slope about 0.3 kg/m. The horizontal data scattering is due to the uncertainty on $\langle v^3 \rangle$ obtained from four identical experimental runs. } 
\label{Ivsv3} 
\end{figure}

Previous studies predict a power law decay proportional to $f^{\gamma}$ with $\gamma \approx 0.6$ followed by an exponential cut-off ~\cite{Mordant,Cadot}. We do not observe such a universal power regime in our experiments. 
The discrepancy might be attributed to the higher forcing frequencies used in our experiment. This specific shape of the spectrum and the value of $\alpha$ is discussed in the next section. Despite the lack of a power law cascade, we  search for a signature of turbulence behavior by studying the scaling laws of the injected power $I$. More precisely, a signature of nonlinear energy transfer is that the the energy flux $\epsilon\propto \langle I\rangle$ is proportional to the third moment of the $v$ \cite{Humbert,Connaughton}. We confirm this scaling on figure \ref{Ivsv3}. By comparing the range of the x-axis of figures \ref{tauvsv} and \ref{Ivsv3}, we notice that despite being small, the skewness of the velocity field, $\langle v^3 \rangle/\sigma_v^3$, is significantly non-zero. This minute deviation from gaussianity is a necessary assumption of the WWT theoretical framework \cite{ZakharovLvovFalkovich,NewellRumpf,Nazarenko}. 

At frequencies lower than the forcing, i.e. $f<f_o$, the main panel of figure \ref{PSDv300} shows a constant spectrum down to a cutoff frequency $f_c^\mathrm{obs.}\sim 30$\,Hz. This observed cutoff occurs at a frequency larger than our estimate based on the plate size $L$, which is $f_c(L)\sim 5$\,Hz. However, it corresponds to a wavelength of nearly half the plate width $f_c(L/2)\sim20$\,Hz. It is not surprising that such low frequencies are affected by the sparse distribution of eigenmodes. We do not observe a dependency of $f_c^{\mathrm{obs.}}$  with the forcing. This supports the assumption that this low frequency cutoff is due to finite size effects \cite{Benjamin2013}. To sum-up our observation, for all forcing parameters explored in our experiment such that nonlinear waves are generated, we obtain the following picture: the temporal spectrum exhibits an  exponential decay for $f>f_o$ and a plateau between $f_c$ and $f_o$. The level of the plateau (shown by the continuous line on figure \ref{PSDv300}) does depend on the forcing parameters, though. 

{\bf Discussions}

We first address the question of the exponential decay of the PDS by means of simple dimensional analysis \cite{Mordant,Connaughton}. The dimensionless parameters are:\begin{gather}
\widetilde{S}_v=S_v/(ch), \quad \phi=fh/\epsilon^{1/3},\\ \phi_o=f_oh/\epsilon^{1/3}, \quad \xi=\epsilon/c^3.\end{gather}
 We recall that $c$ is a characteristic velocity that depends on the material properties, and we denote $f^{\star}=\epsilon^{1/3}/h$ the rescaling frequency previously introduced in \cite{Mordant,Cadot}. $\epsilon$ (in m$^3$/s$^3$) is the mean energy flux by unit of mass injected in the 2D waves. It is proportional to the mean injected power and quantifies the forcing intensity. We deliberately omit from this list the aspect ratio $\Gamma=h/L$ which has been kept constant. Following dimensional analysis, we seek a relation of the form $\widetilde{S_v}=F(\phi,\phi_o,\xi)$, where $F$ is an unknown function. We infer from figure~\ref{PSDv300} that  $F(\phi,\phi_o,\xi)$  reduces to $\widetilde{C}(\xi,\phi_o
)\exp(-\tilde{\alpha}\phi)$  for all $f\geq f_o$ in the limit $f_o\gg f^{\star}$. Note that the characteristic length extracted from figure~\ref{tauvsv}, $\alpha\simeq 0.35$ mm, is of order of $h=0.5$ mm the thickness of the plate.

Secondly, we focus on the low frequency plateau using some scaling arguments. Energy is the unique conserved quantity during nonlinear interactions of bending waves \cite{DuringJosserandRica}. In contrast to surface gravity waves in fluids, there is no other quantities (as wave action) to sustain an inverse cascade. As a consequence, in the stationary regime, one assumes that there is no net energy flux through the lowest wavenumbers, which contribute marginally to the overall dissipation \cite{Lvov}. It is therefore natural to expect equipartition of energy for these modes. With our data and the empirical law for the dissipation presented in \cite{Humbert}, we confirm that at most less than 10\,\% of the power is dissipated below $f_o$ (in the worst case i.e for  $f_o$= 300 Hz).  In the following, we will assume at first approximation that this has no influence on the low frequencies and does not perturb the measured plateau beyond the experimental uncertainty.

In order to explain the plateau, we now adapt the arguments exposed in \cite{GuillaumePRL} to the present case. We start with the isotropic energy spectral density per unit density and per unit surface $e(k)=|\widehat{v}(k)|^2$ where $\widehat{v}(k)$ is the Fourier transform of  velocity. The equipartition of energy implies in a continuous limit  (i.e. $k\gg 1/L$) : 
\begin{equation}
e(k)dk=\frac{1}{\beta}\frac{2\pi k dk}{(2\pi/L)^2}\frac{1}{\rho L^2}\Rightarrow e(k)=\frac{k}{2\pi\rho\beta}.
\label{equipart}
\end{equation}
where $\rho$ is the stainless steel density and $\beta^{-1}$ is the typical energy per modes (analogous to a temperature). 
 The power density spectrum of the perpendicular velocity is simply: $S_v(k)=e(k)/L$. Using the power density spectra relation $S_v(k)dk=S_v(f)df$ together with the dispersion relation, one gets:
\begin{equation}
S_v(f)=\frac{1}{2\beta\rho c L h}.
\label{Svel}
\end{equation}
In other words, the equipartition of energy corresponds to a flat frequency-spectrum in agreement with the experimental spectrum at low frequency.

\begin{figure}
\vspace{-4cm}
\centerline{\resizebox{0.55\textwidth}{!}{%
\includegraphics{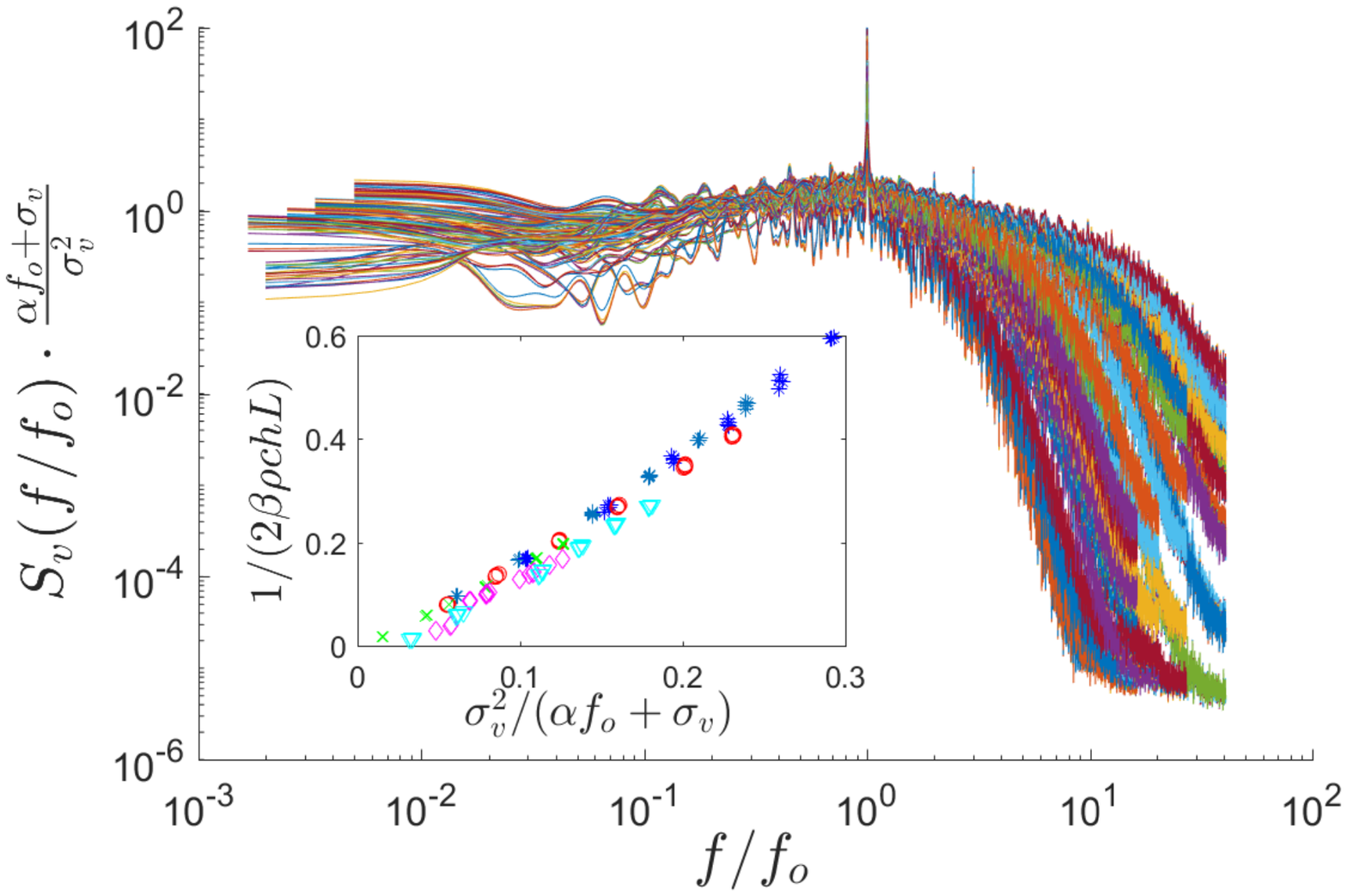}}
}
\vspace{-4cm}
\caption{PDS of $v$, the velocity of the perpendicular displacement of the plate, rescaled by the expected plateau value: $\sigma_v^2/(\alpha f_o+\sigma_v)$ as a function of the reduced frequencies $f/f_o$. Inset: linear relation between the plateau value and $\sigma_v^2/(\alpha f_o+\sigma_v)$ with the forcing frequency: $f_o=100$ Hz  $*$(blue online), $f_o=150$ Hz $\circ$ (red) , $f_o=200$ Hz $\times$ (green) , $f_o=250$ Hz $\diamond$ (magenta), $f_o=300$ Hz $\nabla$ (cyan).}  
\label{Final} 
\end{figure}

Next we determine the variations of $\beta$ with the driving parameters. As observations suggest, we match the low-frequency plateau and the direct cascade at the forcing frequency.
We decompose the PDS into:
\begin{align}
S_v^<(f)&=\frac{1}{2\beta\rho c L h} \text{ for } f\leq f_o\, , \\
S_v^>(f)&=C\exp(-\tau_o f) \text{ for } f>f_o\, ,
\end{align}
with $\tau_o\propto h/\sigma_v$ and where we have actually two unknown $\beta^{-1}$ and $C$. Neglecting the peak and the role of the frequency below $f_c$, in the energy spectrum, one requires that $S_v^<(f_o)=S_v^>(f_o)$ by continuity and that $\sigma_v^2=\int_0^\infty S_v(f) df=\int_0^{f_o} S_v^<(f)df+\int_{f_o}^\infty S_v(f)^>df$. These two relations imply:
\begin{align}
\label{stateEq1}
C&=\frac{\alpha \sigma_v^2}{\alpha f_o+\sigma_v}\exp(\alpha f_o/\sigma_v)\, ,\\
\label{stateEq2}
\beta^{-1}&=2\rho ch L \cdot \frac{\alpha \sigma_v^2}{\alpha f_o+\sigma_v}\, .
\end{align}
Figure~\ref{Final} shows that all the plateaus are well rescaled by: $S_v(f/f_o)\cdot (\alpha f_o+\sigma_v)/ \sigma_v^2$ and that $\beta^{-1}$ is a linear function of $\sigma_v^2/(\alpha f_o+\sigma_v)$. Moreover from figure \ref{Ivsv3} one expects $\sigma_v\sim I^{1/3}$. Thus one writes $\beta^{-1}$ as a function of the forcing parameters: $\beta^{-1}\sim2\rho ch L\cdot (\alpha I^{2/3})(\alpha f_o+ I^{1/3})$. Note that the ratio $\alpha f_o/\sigma_v$ is of order one. Thus, equations (\ref{stateEq1}) and (\ref{stateEq2}) cannot be simplified further in our range of forcing parameters.

{\bf Concluding remarks}

In sum, bending waves forced in a thin elastic plates prove a useful setup to explore the low frequency spectrum of turbulent systems in absence of an inverse cascade. We observe in this system that mean energy is evenly distributed between modes. Finally, we propose an expression allowing a full characterization of the low frequency part  of the velocities spectrum knowing the injected power and the forcing. The build-up of this equipartition possibly occurs through a transient inverse cascade as reported in \cite{DuringJosserandRica}. An experimental study of the transient associated with high frequency forcing would be a challenging endeavor to generalize previous studies of direct cascade transients in elastic plates \cite{MiquelMordantPRL11}. The variations of this mean energy per mode, $\beta^{-1} $, with the forcing parameters is obtained by assuming the continuity of the spectrum. Ideally, one would aim at observing simultaneously energy equipartition for low frequencies $f\le f_0$ and a direct cascade (characterized by a power law spectrum) for $f\ge f_0$, before dissipative effects take over. In the present experiment, high driving frequencies were necessary so that equipartition was observed over a convincingly large range of frequencies. As an unfortunate consequence, the small scales did not exhibit a conservative cascades but were affected by dissipation. A goal for future studies, perhaps achievable using a much larger plate, would be to somehow extend the range of the ``transparency window'' (i.e. the range of scales where energy is conservatively transferred) so that to observe simultaneously equipartition and a direct cascade that is ultimately dissipated at small scales. 

\end{document}